\documentclass[pra,twocolumn]{revtex4}
\usepackage{graphicx}
\usepackage{graphics}
\usepackage{amssymb}
\usepackage{epstopdf}
\usepackage{color}
\usepackage{subfigure}

\begin{document}
\title{Spatial Entanglement of a Free Bosonic Field}
\author{Libby Heaney$^1$, Janet Anders$^2$, Vlatko Vedral$^{1,2}$}
\affiliation{Department of Physics and Astronomy, EC Stoner Building, University of Leeds, Leeds, LS2 9JT, UK$^1$\\
Quantum Information Technology Lab, Department of Physics, National University of Singapore, Singapore 117542 $^2$} 

\date{\today}

\begin{abstract}
In this paper we discuss the entanglement properties of a thermal non-relativistic free bosonic field.   We demonstrate how to formally construct spatial modes in order to use a continuous variable separability criterion and show that the construction of the modes plays a significant role in the way the entanglement manifests itself.
For instance the presence of entanglement depends on the shape of the modes and their separation.  The temperature of the gas is another crucial factor and for one choice of modes we show that entanglement can be found at arbitrarily high temperatures as long as we can zoom in on suitably small regions in space.   Moreover, we show that the entanglement here is useful as it can be extracted to a pair of localised systems.
\end{abstract}

\maketitle

\section{Introduction}

Entanglement, a fundamental ingredient of quantum mechanics, leads to correlations between subsystems that are greater than anything possible classically and is the main resource for quantum computations.  However these quantum correlations are delicate and can easily be destroyed.  Therefore it is logical to hunt out systems that possess an amount of natural thermal entanglement when they are under certain conditions, for instance below a certain temperature, so that if one keeps the conditions stable, the entanglement definitely exists even though the system may be open.

Unlike previous papers that use macroscopic variables to witness entanglement in discrete systems - for instance temperature \cite{hightempent}, heat capacity \cite{heatcapent} and magnetic susceptibility \cite{ghosh,Vedral1}, we will use an existing separability criterion to investigate whether a Bose field is entangled and look at how such variables affect the entanglement. This is useful to know as entanglement found at higher temperatures is easier to obtain than in situations where the exact ground state is vital \cite{Dunning05, Latorre04}.  
 
We will focus on spatial entanglement, which means that localised regions of a system exhibit particle number correlations that are non-local. 
When considering spatial entanglement it is necessary to work in the second quantised formalism, as entanglement in many-body systems can exist between the occupation number of modes and not just between internal degrees of freedom of the particles themselves.  This allows for the interesting case of single particle entanglement, which was elsewhere \cite{vlatkomarcellojacob} shown to be of the same form as the usual EPR entanglement.  
Crucially, the first quantisation is misleading since it can generate inconsistent conclusions.  If we worked in the first quantised picture, defined two spatially separated regions, the Hilbert space would not  be a direct product of the Hilbert spaces of the two regions and cannot be decomposed into spatial modes.  Thus we must work in the second quantised picture where each region represents an independent mode and our Hilbert space is a tensor product of the two.  We will proceed to investigate entanglement of the free bosonic field in this manner.
This is an interesting problem as not only will we show that spatial entanglement can be extracted for computational purposes but the physics here is curious as we find an underlying entangled structure in a relatively simple system.   

To motivate our work further note that a recent paper \cite{anders} provided a general argument for why energy might act as a good witness for spatial entanglement in CV many-body systems and demonstrated that the critical temperature for this entanglement is remarkably similar to the critical temperature for Bose-Einstein condensation.  In the present paper we put foundations to this problem by tackling it from an entirely different direction.  We formally construct spatial modes in terms of ladder operators for regions of space and apply an existing separability criterion \cite{simon} which unambiguously decides whether the two spatial modes are entangled.    Unlike \cite{anders}, in this constructive scenario, it is clear how the entanglement emerges and our criterion is both necessary and sufficient for bi-partite divisions in its most general form.  Additionally, we will discuss the results in terms of thermal states, give an extraction procedure and contrast our methods to another related paper \cite{Narnhofer03} .

The paper is organised as follows.  Let us proceed by firstly providing an overview of Gaussian states and the separability criterion that we will use to detect the entanglement in section \ref{gaussianstatessection}.  Therefore if the reader is familiar with Gaussian states please go directly to section \ref{mainsection}.  We next analyse the spatial entanglement in the gas and discuss the role of the detector profile in section \ref{mainsection}.  Finally we show that the entanglement is useful as it can be extracted to a pair of localised systems in section \ref{extraction}.  
 
 \section{Gaussian States}\label{gaussianstatessection}

Gaussian states are common in the physical arena as they closely approximate coherent and squeezed states of light  \cite{Loudon73} and thermal states of a quadratic Hamiltonian are Gaussian.  The set of Gaussian states have Gaussian characteristic functions and are therefore expressed entirely in terms of their first and second moments.  We will now develop a mathematical formalism to allow us to look at two-mode Gaussian states closer.   Please note that complete reviews of Gaussian states can be found in \cite{Anders03, Braunstein05, Adesso07}.

A quantum bosonic field is equivalent to a (possibly infinite) set of independent quantum harmonic oscillators.  Thus for a known number of bosons the Hilbert space $\mathcal{H}=\bigotimes_{k=1}^{2}\mathcal{H}_k$ of such a system is a tensor product of single-oscillator Fock spaces $\mathcal{H}_k$.   The creation and annihilation operators $\hat{a}_k^{\dagger}$ and $\hat{a}_k$ acting on the Fock spaces can then be related to the phase-space operators through 
\begin{equation}
\label{positionandmomentum}
\hat{u}_k=\sqrt{\frac{\hbar}{2m\omega_k}}(\hat{a}_k^\dagger+\hat{a}_k),\quad\quad\hat{p}_k=i\sqrt{\frac{\hbar m \omega_k}{2}}(\hat{a}_k^\dagger-\hat{a}_k),
\end{equation}
where $\omega_k$ is the angular frequency of the $k$th mode.
If the operators are arranged in a vector as follows
\begin{equation}
\hat{\xi}=(\hat{u}_1,\hat{p}_1,\hat{u}_2,\hat{p}_2),
\end{equation}
the canonical commutation relations can be expressed in the compact form as 
\begin{equation}
\label{compactCCR}
\big{[}\hat{\xi}_i,\hat{\xi}_j\big{]}= i\hbar\, \Omega_{ij},
\end{equation}
where the subscripts $i,j=1,2,3,4$ refer to the different components of this vector.  The matrix $\Omega_{ij}$ is defined as
\begin{equation}
\Omega\equiv \bigoplus_{k=1}^2J,\quad\quad J\equiv\left(
\begin{array}{cc}
0 & 1\\
-1 & 0
\end{array}
\right).
\end{equation}

It is well known that the state of the gas is described by the positive trace-class density operator $\rho$ that is used to calculate expectation values of the phase-space operators, which form the first and second moments describing the Gaussian state.  These can be grouped together in the covariance matrix (CM), which is defined through its elements as
\begin{equation}
\label{CMbyelements}
\gamma_{ij}=\langle[\hat{\xi}_i,\hat{\xi}_j]_{+}\rangle-\langle\hat{\xi}_i\rangle\langle\hat{\xi}_j\rangle,
\end{equation}
where $\langle\hat{O}\rangle = \textrm{Tr}[\hat{O}\rho]$.
The CM $\gamma$ is real, symmetric and holds all the available information about the state of the gas and like the density operator $\rho$ the CM must fulfill certain criteria to be a {\it bona-fide} CM.  
Note that the displacements $\langle\hat{\xi}_{i(j)}\rangle$ can be altered by local operations and classical communications (LOCC) and do not affect the entanglement properties of the system and are set to zero here without loss of generality.  

Thus, using (\ref{compactCCR}) and (\ref{CMbyelements}) is it straight forward to show that the CM must obey following inequality 
\begin{equation}
\label{compactuncertainties}
\gamma+i\Omega\geq 0.
\end{equation}
This inequality is the only requirement for the CM of the gas to represent a physical state.

It will be useful to manipulate the CM in order to simplify the calculations. Using local symplectic transformations $S_1$ and $S_2$ such that $S=S_1\oplus S_2$ preserve any entanglement properties of the CM, a ten parameter CM of two modes can be reduced to only four parameters.  The CM is now in its standard form $\gamma^{SF}$ and is expressed as
\begin{equation}
\label{standardform}
S^T\gamma S=\gamma^{SF}=\left(
\begin{array}{cccc}
a & 0 & c & 0\\
0 & a & 0 & d\\
c & 0 & b & 0\\
0 & d & 0 & b
\end{array}
\right).
\end{equation}
There are four symplectic invariants associated with the CM, namely $\mathrm{det}\gamma=(ab-c^2)(ab-d^2), \mathrm{det}A=a^2,\mathrm{det}B=b^2$ and $ \mathrm{det}C=cd$.  These invariants are unique up to the relative sign of $c$ and $d$ for any covariance matrix $\gamma$.  The symplectic eigenvalues \cite{Anders03,Adesso07} are defined as the ones for $\gamma$ itself and are expressed in terms of the invariant quantities ($a,b, ... etc.$) as
\begin{equation}
\label{symplecticeigenvalues}
\nu_{\pm}=\frac{1}{\sqrt{2}}\sqrt{a^2+b^2+2cd\pm\sqrt{(a^2+b^2+2cd)^2-4\mathrm{det}\gamma}}.  
\end{equation}
The uncertainty relations on the CM are satisfied if $\nu_{-}\geq1$.  It is known that only pure states can saturate this bound while mixed states necessarily remain above it. 

When dealing with entanglement it is often useful to calculate the purity of the state under consideration because as a states purity increases its entanglement usually increases too .  When using a density operator $\rho$ we can calculate the purity through $\mu=\textrm{Tr}[\rho^2]$, which translates into
\begin{equation}
\label{purity}
\mu=\textrm{Tr}[\rho^2]=\frac{1}{\sqrt{\textrm{det}\,\gamma}},
\end{equation}
where for pure states $\textrm{det}\,\gamma=1$ and $\textrm{det}\,\gamma>1$ for mixed states.

To investigate entanglement of a two-mode Gaussian state we use a criterion by Simon \cite{simon} that defines the partial transpose operation for CV states.  The transposition operator used to detect entanglement in \cite{Peres, Horodecki} is equivalent to partial complex conjugation (or {\it time reversal}) of a CV system.  Simon saw this as a 'mirror reflection in phase space',  however, a mirror reflection in phase space is not a canonical transformation and cannot be physically implemented. It is used here as a mathematical way of detecting entanglement. 

To apply this operation to a bi-partite CV state we change the sign of the momenta of one constituent subsystem.  As in \cite{Peres, Horodecki} when the transposition is applied to one part of an entangled system it causes the state to become unphysical, the uncertainty relations (\ref{compactuncertainties}) with our new CM, $\gamma^{T_A}$ will become unphysical if the state is entangled.  In fact, it is straightforward to check this relation using the symplectic eigenvalues (\ref{symplecticeigenvalues}) and $\nu_{-}\geq1$.

In addition it is useful to know the amount of entanglement in a system, which can be calculated through the logarithmic negativity \cite{lognegativity} 
\begin{equation}
E_\mathcal{N}=\textrm{max}\{0,-\log_2(\nu_-)\}.
\end{equation}
The logarithmic negativity $E_\mathcal{N}$ is therefore zero for separable states and increases unboundedly as $\nu_-\rightarrow 0$.

\section{Entanglement Between Spatial Modes}\label{mainsection}

The formalism in the previous section allows one to calculate the entanglement between two regions of a field, but first, the regions need to be constructed so that the separability criterion from the previous section is applicable.  

\subsection{Defining Spatial Modes}
 
\begin{figure}[]\centering
{\includegraphics[width=0.35\textwidth]{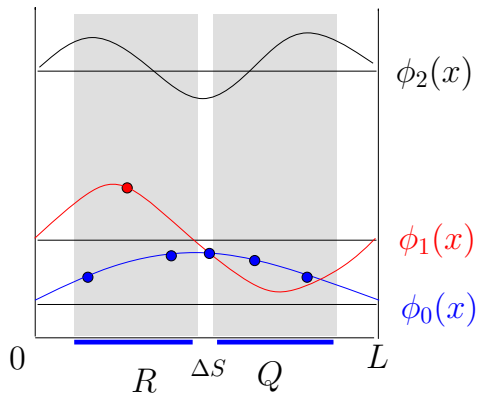}
\caption{Visualisation of the system, in which a thermal non-interacting bosonic gas is trapped in a 1D infinite square well of length $L$ and the particles occupy decoupled momentum modes $\phi_k(x)$.  We wish to investigate entanglement between two spatial modes $R$ and $Q$ (grey shading) with a finite distance between them, denoted by $\Delta S$. }\label{regions}}
\end{figure}

Before we construct the spatial modes (regions) themselves we shall briefly describe in more detail the system under consideration.  For simplicity we take a 1D infinite square well  potential of length $L$, see Fig. \ref{regions}, filled with a free bosonic gas in a thermal state $\rho = \exp(-(\hat{H}-\mu \hat{N})/k_BT)/Z$ at temperature $T$. The quantities $\hat{H}$, $k_B$, $\mu$ and $\hat{N}$ are the Hamiltonian,  Boltzmann's constant, the chemical potential and the number of particles in the system, respectively.    The grand partition function is $Z=tr[\exp(-\beta(\hat{H}-\mu\hat{N})]$. Note, that confining geometries of higher dimensions could be considered but this does not contirbute much to the present evaluation.
In momentum representation, the Hamiltonian of a non-interacting bosonic gas in its second quantised form is simply $\hat{H}=\sum_k E_k \hat{a}_k^{\dagger}\hat{a}_k$,
where $k$ are the momentum modes of the system , $\hat{a}_k$ and $\hat{a}_k^{\dagger}$ are the annihilation and creation operators of particles in these modes and $E_k = \hbar\omega_k$ with the usual dispersion relation for massive particles $\omega_k=\hbar k^2 /2m$.  As the Hamiltonian is quadratic the thermal state is Gaussian and can be described entirely by its first and second moments.  

The  Hamiltonian describes the occupation of decoupled momentum modes $\phi_{k}(x)=\sqrt{2/L}\,\sin(k x)$ with $k=\pi l/L$ and consequently the state $\rho$ is separable w.r.t. a decomposition into momentum modes.  Nevertheless entanglement could exist in position space and we wish to investigate any spatial correlations.   To apply the CV separability criterion \cite{simon},  position and momentum operators for the regions should be constructed.  In lattice systems the idea of spatial modes is clearly defined in terms of individual or groups of lattice sites and the corresponding operators would be the position and momentum of the particles at the lattice sites.  However, in a CV system one has the freedom to include any region of space in the spatial mode, which may in turn contain any number of particles.  The modes are then constructed by averaging position and momentum operators with a complex detector profile over the set of points belonging to the region.  Although our operators must be position and momentum, they will therefore be the averaged position and momentum of all the particles (or excitations) in the spatial modes and not the position and momentum of the particles themselves.  

In fact, position and momentum operators acting on Fock spaces of individual momentum modes were defined in the previous section (\ref{positionandmomentum}).   So, in order to construct the spatial modes here we firstly take those phase-space operators (\ref{positionandmomentum}) and transform into position representation with the discrete Fourier transform $\hat{u}(x)=\sum_{k=1}^{\infty}\phi_{k}(x)\,\hat{u}_{k}$ and similar for $\hat{p}(x)$.    And secondly, as space - unlike momentum in our system - is continuous and a point in space is of zero volume and hence meaningless to talk about, we investigate entanglement between spatial regions of a finite size, denoted by $R$ and $Q$ (Fig. \ref{regions}).  The conjugate operators $\hat{u}(x), \hat{p}(x)$ become $\hat{u}_R, \hat{p}_R, \hat{u}_Q$ and $\hat{p}_Q$ by averaging over a detector profile $g(x)$ localised in $R$ and $Q$, for example the operator $\hat{u}_R$ is $\hat{u}_R = \int_R dx \,g(x)\hat{u}(x)$.  

We can take any normalised complex function for $g(x)$ and of course there are an infinite number to choose from. If we find just one $g(x)$ for which the separability condition is violated,  the gas is spatially entangled.  On the other hand, the gas is only truly separable if there is no detector profile that gives entanglement, but to confirm this one would have to search over infinitely many $g(x)$.  Naturally our operators must obey the usual commutation relations $[\hat{u}_R, \hat{p}_Q]=[\hat{u}_R,\hat{u}_Q]=[\hat{p}_R,\hat{p}_Q]=0$ and $[\hat{u}_R,\hat{p}_R]=i\hbar$, which determine the normalisation of $g(x)$.  

Now that we have constructed the operators that define our spatial regions, we can apply the separability criterion \cite{simon} to the gas.

\subsection{Applying The Separability Criterion}

It is straight-forward to check that the transformation between the momentum and spatial modes is linear and the Gaussian nature of the state in the new canonical operators is preserved.  This means that the state can be described fully by a two-mode CM (\ref{CMbyelements}), with the two modes representing regions $R$ and $Q$. 

By expanding the operators $\hat{\xi}_{i}$ in terms of the reciprocal space annihilation and creation operators, $\hat{a}_k$ and $\hat{a}_k^{\dagger}$, the constituents of the covariance matrix can be evaluated easily in the momentum Fock basis.  The resulting CM is of the form
{\begin{equation}
\label{abrevgamma}
\gamma =
\left( \begin{array}{cccc}
A & 0 & E & 0\\
0 & B & 0 & F\\
E & 0 & C & 0\\
0 & F & 0 & D
\end{array} \right).
\end{equation}}

For completeness the general form of the CM elements are
\begin{eqnarray}
\label{elements}
A,C = 2\langle \hat{u}_{R}^2\rangle_{\rho} &=& 2 \textrm{tr} \Big{[} \Big{(} \int_{R,Q} dx \,g(x) \hat{u}(x)\Big{)}^2\Big{]}\\
&=&   \sum_k \frac{1}{\pi^2k^2}(\mathcal{I}_{R,Q}^k)^2 \coth\Big{(}\frac{(E_k-\mu) \beta}{2}\bigg{)}\nonumber\\
B,D=2\langle \hat{p}_{R,Q}^2\rangle_{\rho} &=&  \sum_k \pi^2k^2 (\mathcal{I}_{R,Q}^k)^2 \coth\bigg{(}\frac{(E_k-\mu) \beta}{2}\bigg{)}\nonumber\\
E = \langle \{\hat{u}_R,\hat{u}_Q\}\rangle &=&  \sum_k \frac{1}{k^2\pi^2}\mathcal{I}_{R}^{k}\mathcal{I}_{Q}^{k}\coth\bigg{(}\frac{(E_k-\mu)\beta}{2}\bigg{)}\nonumber\\ 
F = \langle \{\hat{p}_R,\hat{p}_Q\}\rangle &=& \sum_k \pi^2 k^2\mathcal{I}_{R}^{k}\mathcal{I}_{Q}^{k}\coth\bigg{(}\frac{(E_k-\mu)\beta}{2}\bigg{)}\nonumber,
\end{eqnarray} 
where $\mathcal{I}_{R,Q}^{k}=\int_{R,Q}dx\,g(x)\phi_k(x)$.
The factor $\sqrt{\hbar/2m\omega_k}$ in the operator $\hat{u}_k$ results in the $1/\pi^2k^2$ in $A,C$ and $E$ above (\ref{elements}) and terms containing $k^2$ occur in the sums for the momenta uncertainties (elements $B$, $D$ and $F$). These terms play an important role in the convergence/divergence properties of the elements and consequently in the resulting entanglement. 

Everything is now expressed suitably for the CV separability criterion described in section \ref{gaussianstatessection}.
We would like to arrive at an inequality that must be positive for separable states and negative for entangled states so let us apply the partial transposition which means $F\rightarrow -F$.   Examining the symplectic eigenvalue $\nu_-^{T_A}$ of the partially transposed CM leads to the following expression in terms of the original CM elements,
\begin{equation}
\label{sepcondition}
\mathcal{E} = 1 + (AC - F^2)(BD - E^2) - AB - CD+2EF.
\end{equation}
If $\mathcal{E}\geq0$ the two spatial modes are separable, on the other hand $\mathcal{E}<0$ means that the modes are entangled.  

However, before we can even speak about entanglement between the spatial modes the CM must represent a physical state, which is the case when the uncertainty relation (\ref{compactuncertainties}) for the original CM $\gamma$ is satisfied.

\subsection{Choosing A Detector Profile}

Before we can check the entanglement of the gas through (\ref{sepcondition}), a specific form for the so-called detector profile $g(x)$ must be assumed.  In other words how should the set of points belonging to the regions $R$ and $Q$ be averaged over?  

\subsubsection{Top Hat Detector Profile}

Let us start simply by giving equal weight to all the points belonging to the regions i.e. we use a top-hat function that gives support inside the regions and is zero outside of them.   This is a natural choice as in practice many detectors click if something is in a certain range but will not weight any two positions in that range differently.  

With this choice of detector profile the system is always separable w.r.t. the two spatial modes, independently of the size of the regions, their separation and the temperature.  
This is intuitively clear as in a thermal state all possible momenta are mixed and correlations are lost.  Mathematically this is clear as the momentum entries $\langle \hat{P}_R^2\rangle$ etc  in our CM contain terms proportional to $k^2$ in the sums over $k$ and will not converge.  To combat this and arrive at a less mixed state, the state could be prepared so that it contains only a particular portion of the mixture.  However entanglement can be found without manipulating the state at all, so let us look firstly at this.  Although later we will come back to using a top hat function as the detector profile and allow for some selection of momenta (see section \ref{tophatstateprep}), because interestingly in that case entanglement is found to exist, under some circumstances, at very high temperatures.

\subsubsection{Gaussian Detector Profile}\label{gaussianprofile}

In the previous section the detector profile $g(x)$ was chosen to be a top hat function, however this meant that all momenta were detected with equal weighting and no entanglement was uncovered.   Conversely, one could have chosen $g(x)$ so that it naturally gave a lower weighting to higher momentum modes so that the state becomes less mixed.  This is what we will try next.    

The freedom to choose the detector profile illustrates that there are  infinitely many choices of modes between which there could be entanglement.  This is because a free gas has infinitely many degrees of freedom.  If we find one pair of spatial modes that are entangled than we can safely say that the gas is entangled. One could choose any detector profile, but by fixing the detector profile the entanglement between any other combinations of modes is excluded and is therefore less than the actual amount of entanglement in the gas. 

Naturally it is difficult to find the  $g(x)$ that picks up the entanglement, so instead we start by taking the function $\mathcal{I}_{R,Q}^{k}=\int_{R,Q}dx\,g(x)\phi_k(x)$, which occurs in (\ref{elements}) and see that for the top hat detector profile it was $\mathcal{I}_{R,Q}^{k}=\textrm{C}/k\sqrt{2/L}(\cos(R_1(Q_1)k/L)-\cos(R_2(Q_2)k/L))$, where $R_1(Q_1)$ and $R_2(Q_2)$ denote the edges of the regions $R$ or $Q$.  We use this as the starting point to define a new $\mathcal{I}_{R,Q}^{k}$ and combine it with a function that gives greater weighting to a set of low momenta depending on the size of the region under consideration.  Thus $\mathcal{I}_{R,Q}^{k}=\textrm{C}(\cos(R_1(Q_1)k/L)-\cos(R_2(Q_2)k/L))\exp(-k^2\Delta x^2)$, where the normalisation C is determined through the commutation relations. Here, although the detector profile has not been specified directly, we have chosen one that influences the range of momenta it picks up $\Delta k$ depending on the size of the region $\Delta x$.  

The new choice of detector profile has the desired effect as entanglement is now found in the gas without any other manipulations.

\begin{figure}[]\centering
{\includegraphics[width=0.5\textwidth]{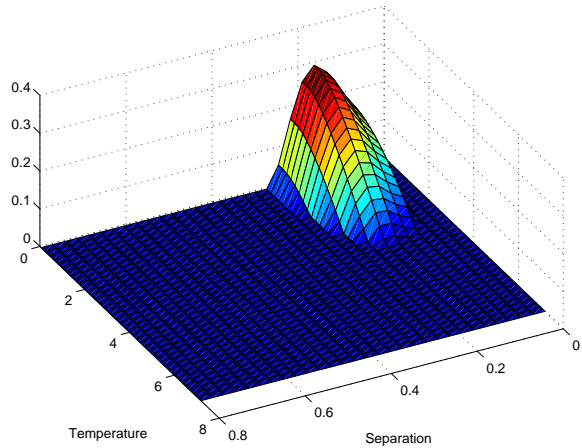}}
\caption{The plot shows the logarithmic negativity for the region size $\Delta x=L/10$ against the separation of the regions and the temperature of the gas.  The amount  of entanglement increases as the temperature is lowered and only exists when the regions are sufficiently close to one another.   \label{logneg}}
\end{figure}

Let us look now at how the entanglement manifests itself.  Firstly, unlike for a top hat detector profile if momenta in the state were  pre-selected (as discussed in the following section \ref{tophatstateprep}), here the entanglement cannot exist at arbitrarily high temperatures even if the regions are very small as in this case the uncertainty relations (\ref{compactuncertainties}) are not satisfied.  Entanglement exists for regions in the size range of $\Delta x \sim L/20 \rightarrow L/2$ and the amount of entanglement for regions of the size $\Delta x=L/10$  and for non-zero temperatures of the gas is plotted in Fig. \ref{logneg} using the logarithmic negativity  as a function of their separation.  There is no entanglement above a certain temperature and when the regions are far away from one another.  

\subsubsection{Top Hat Detector Profile and Selecting Momenta}\label{tophatstateprep}

It is clear that a thermal Bosonic field is entangled when a suitable detector profile that limits the mixedness of the gas is chosen. In order to compare the present work with \cite{anders} and illustrate the interesting case of entanglement at very high temperatures, we would like to show how one can still find entanglement in the gas with top hat detector profile by preparing the state with the necessary mixedness.  

It is straight-forward to examine the mixedness of the gas using the purity relation given in eq. (\ref{purity}) as the entanglement criterion (\ref{sepcondition}) can be written in terms of the purity
\begin{equation}
\label{purityineq}
\mu>\frac{1}{\sqrt{AB+CD-2EF-1}}.
\end{equation}
The gas can be prepared with a greater purity by projecting out all configurations over a certain momenta.  Once the purity satisfies (\ref{purityineq}) the gas will be entangled.   However, with a top hat detector profile this action cannot be performed locally as we cannot distinguish the momentum modes in the spatial regions alone as they are only orthonormal w.r.t. the entire well.  The projection is therefore a global operation, but this is just part of the state preparation as we are defining the shape of the spatial modes upon which the entanglement very much depends.  Indeed, the projection proves to be a profitable exercise as afterwards $\mathcal{E}<0$ and we uncover spatial entanglement, under certain circumstances, between the two spatial modes. 

\begin{figure}[t]\centering
{\includegraphics[width=0.4\textwidth]{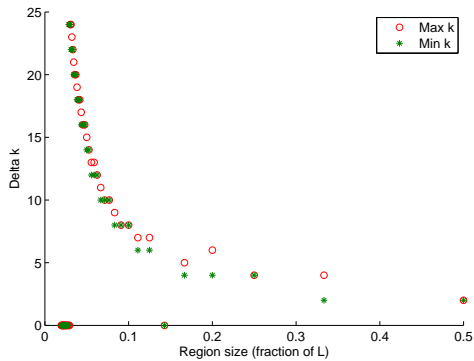}
\caption{We plot the number of momenta ($\Delta k$) that is included in an entangled state against the size of the regions $\Delta x_R=\Delta{x_Q}=\Delta x$ when the two regions are touching at $L/2$.  We sum from the lowest momentum $k=\pi/L$ up to $k_{min}$ where one first finds entangled states (this is our $\Delta k_\textrm{min}$).  For some $\Delta x$, entangled states include a wider range of momenta from $k=\pi/L$ up to $k_{max}$ ($\Delta k_\textrm{max}$).  It is evident that for smaller regions entangled states must include an increasing number of modes in order to satisfy the uncertainty relations.  The shape of this graph illustrates that $\Delta k \sim 1/\Delta x$ is obeyed.}\label{DeltakVsReg}}
\end{figure}

We know that for entanglement to emerge, the gas has to be of a certain purity which means that the state must only include particles below a given momenta.  But how does one estimate the maximum momenta that an entangled field should have?  It turns out that the maximum momenta for an entangled set of particles is determined by Heisenberg's uncertainty relations.  It was noted in the previous section that the CM must fulfill the uncertainty relation (\ref{compactuncertainties}) if it is to represent a physical state. In simple terms, this means that the state should occupy an area in phase-space greater than $\Delta x\Delta k\sim1$, where $\Delta x$ is the size of the regions and $\Delta k$ is the number of different momenta in the state. When defining the spatial regions we have limited the size of our $\Delta x$ so consequently we must ensure that the state is prepared with an adequate $\Delta k$ to satisfy this relation and that we do not try to measure both quadratures with a greater precision than possible.  However, only by considering states that have the smallest possible uncertainty is the entanglement uncovered (Fig. \ref{DeltakVsReg}).  Fig. \ref{DeltakVsReg} indicates how many momenta have to be included in our state for the two regions to be entangled at a fixed $T$ and confirms that $\Delta k\sim1/\Delta x$ is roughly obeyed.

\begin{figure}[t]
{{\includegraphics[width=0.4\textwidth]{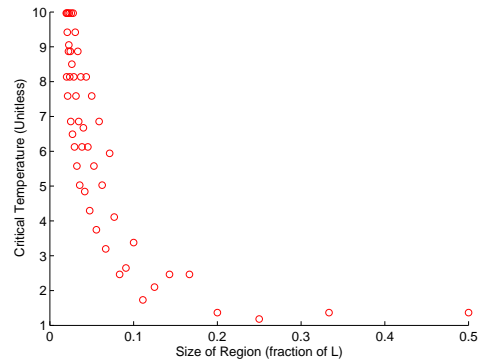}}
\caption{The critical temperature $T_C$ as a function of the size of the regions, where $\Delta x_R=\Delta x_Q$ and the regions are touching at $L/2$.  For larger regions the critical temperature is very low, but for smaller  and smaller regions $T_C$ increases. }\label{SizeTemp}}
\end{figure}

The entanglement depends upon the width of the regions, their separation and the temperature of the system.  Entanglement for given sized regions exists when the regions are suitably close and  below a certain critical temperature $T_C$.  If we investigate the nature of this critical temperature with respect to the size of the regions, from Fig. \ref{SizeTemp} it is evident that the smaller the regions, the higher the critical temperature for entanglement.  In fact we assert that at very small length scales entanglement exists at arbitrarily high temperatures as the critical temperature behaves roughly as $T_C\propto \Delta x^{-\frac{3}{4}}$ with the size of the regions $\Delta x$.  The high temperature entanglement occurs between two very small regions with a top hat detector profile precisely because of the uncertainty relations (\ref{compactuncertainties}).  On one hand, the Gaussian detector profile suppressed the occupation of higher momentum modes at high temperatures which meant the uncertainty relations were not satisfied.   Whereas here it is always possible to satisfy the uncertainty relations, but in order to generate a $\Delta k$ large enough to do so, the gas has to be heated up to a finite $T$, which results in high temperature entanglement. 

Regardless of the choice of detector profile any spatial entanglement in the gas is in the form of particle number correlations between the two spatial modes and as it is not between the particles themselves, one may question whether the entanglement is genuine and useful or whether it is created by artificially defining the regions.  The following section demonstrates that this is genuine entanglement by describing an extraction process whereby the entanglement is transfered to two localised systems which could then be used for teleportation.

\section{Extracting Entanglement From A Bosonic Gas}\label{extraction}

We know that the entanglement in the gas is strongly dependent on the choice of detector profile $g(x)$ that defines the shape of our region. However, in this section the detector profiles will result from the effective volume of two localised systems (here we take atoms but we could of course take quantum dots or any other viable system), which will interact locally with the gas.  Entanglement will be extracted from the gas so not only will we have proper valuable EPR-type entanglement, but the extraction also provides another way of locally reducing the number of momenta in our state.  

The method we use for extraction is similar to a generalised scheme by Kaszlikowski {\it et al.} \cite{dagoextract} although here we use a different Hamiltonian.  Our set up is as follows.  Two systems - held by Alice ($S_A$) and Bob ($S_B$) - are placed in close proximity to the gas and their natural finite width is used to define the shape (detector profile $g(x)$) of two separate spatial regions.  The systems are localised in two distinct regions of space and start in a separable state with respect to one another.   The probes will interact with the gas with a sufficiently short time in order to deny them from sharing information with one another via the bosonic field. 

The Hamiltonian that provides this interaction is $H(t)=\Gamma(t)(\hat{u}_R\hat{P}_{S_A}+\hat{u}_Q\hat{P}_{S_B})$, where $\Gamma(t)$ is the interaction strength.  The position of the gas located in region $R$ couples uniquely with the momentum of Alice's system $\hat{P}_{S_A}$ and the position of the gas in the other region, $Q$, couples locally to the momentum of Bob's system $\hat{P}_{S_B}$.  To select the entangled set of momenta the probe atoms should have a gaussian shape to them which mimics the $g(x)$ used in the previous section.  This would be the case for instance if an atom was in the lowest energy level in a harmonic trap.

To show that the probe systems are entangled after we have applied the interaction for some time $t$,  we must determine the time evolution of the state.  The system is initially in the state $\rho(0)=\rho_g|00\rangle\langle00|$ (where $|00\rangle=|0\rangle_A\otimes|0\rangle_B$ and here $\rho_{g}$ is the density operator of the bosonic gas) and as the full time evolution is difficult to establish, as we have chosen a suitably short interaction, we can calculate the state at time $\delta t$, perturbatively, to first order.  The state of the gas and probes as  $\textrm{lim}_{\delta t\rightarrow 0}\Gamma=\int_0^{\delta t} dt\, \Gamma(t) <<1$ is $\rho(\delta t\rightarrow0) = \rho(0) + \Gamma^2\hbar m \omega /2(\hat{u}_R\rho_{g}\hat{u}_R|10\rangle \langle 10|+\hat{u}_Q\rho_{g}\hat{u}_Q|01\rangle\langle01| +(\hat{u}_R\rho_{g}\hat{u}_Q|10\rangle\langle01| + \mathrm{h.c.}))+\Gamma^4\hat{u}_R\hat{u}_Q\rho_g\hat{u}_R\hat{u}_Q|11\rangle\langle11|$.

We would like to know whether the probe systems have now become entangled in momentum and for this we trace out the gas and compute the Peres-Horodecki criterion \cite{Peres, Horodecki}.  Before we apply the partial transposition the density matrix of the atoms, written in the basis of the lowest two momentum states, is of the form
\begin{displaymath}
\rho_{Atoms} =
\left( \begin{array}{cccc}
1 & 0 & 0 & 0\\
0 & x & y & 0\\
0 & y & z & 0\\
0 & 0 & 0 & \delta
\end{array}
\right),
\end{displaymath}
where $x\propto$ tr$[\rho\hat{u}_R^2], z\propto$ tr$[\rho\hat{u}_Q^2]$, $y\propto$ tr$[\rho\hat{u}_R\hat{u}_Q]$ and the term $\delta\propto$ tr$[\rho\hat{u}_R^2\hat{u}_Q^2]$ is very small (as it its proportional to $\Gamma^4$) but non-negligible.   If $y>1/2\sqrt{\delta(\delta+1)}$ is satisfied the atoms will become entangled and as $\delta\ll1$ we can approximate this by $y>0$.  If the off-diagonal element $y$ exists then we can extract entanglement from the gas.  The term $y\propto$tr$[\rho\hat{u}_R\hat{u}_Q]$ is the overlap between the positions of regions $R$ and $Q$ and its existence depends upon $g(x)$ the temperature of the gas and the separation of the two regions.  

\section{Discussion and Conclusions}\label{discussion}

Let us now compare our findings to two other papers that investigated field theoretic entanglement. 

Firstly we found that entangled fields had a maximum energy. If we compare this to the results of \cite{anders}, it is not surprising that the minimum energy of the separable set of states is fixed by the uncertainty relation.  There, the authors used the uncertainty relations applied to regions of space to find the minimum energy that separable states can have and said that states with a lower energy are entangled.   Unlike \cite{anders}, here we have tackled the problem constructively by defining proper spatial modes and applying a well known separability criterion and as a result we can see clearly that by restricting ourselves to slow moving particles in the two regions the state becomes purer and therefore becomes entangled.

Secondly, another paper \cite{Narnhofer03} that analyses entanglement of a field, constructs spatial modes quite differently to us.  In \cite{Narnhofer03} each region of space was mathematically converted to a two level system and the field was projected onto these two states thereby reducing an infinite dimension system down to just two degrees of freedom, which allowed the application of the Peres-Horodecki criterion \cite{Peres,Horodecki}.  On the other hand we reduce our field to a harmonic oscillator for each region and allow each one to be populated by bosons.    The latter is more operational than \cite{Narnhofer03} as the way we average over the set of points in each region (i.e. we must specify a detector profile for each region) proves to be crucial for revealing entanglement.

In this paper we have shown that the remarkably simple system of non-interacting bosons is entangled in space.  Unlike discrete systems such as spin chains that have their spatial modes naturally defined here one must create spatial modes by averaging operators over a set of points with some detector profile $g(x)$.  It is therefore natural that the choice of $g(x)$ effects the entanglement that is uncovered.  Thus the entanglement resulting from two different choices of spatial modes has been analysed.  Firstly, if one averages over all points in the regions equally (i.e. the detector profile is a top hat function),  to find entanglement one has to prepare the gas by lowering the number of momenta in the state using a global operation.  In this case however, entanglement exists at very high temperatures if we consider very small regions due to the uncertainty relations.  Secondly, a Gaussian type detector profile that weighted the higher momentum modes less than the lower ones uncovered entanglement locally.  In this case, the entanglement always existed below a certain temperature and disappeared altogether for very small regions. Moreover we have demonstrated that the entanglement is useful as it can be extracted from the gas.

\section{Acknowledgements}

VV and LH acknowledge the support of the Engineering and Physical Sciences Research Council and the Royal Society and Wolfson Foundation in UK for funding and the National University of Singapore for their hospitality.  JA acknowledges support of the Gottlieb Daimler und Karl Benz-Stiftung.  We all thank D. Kaszlikowski for interesting discussions about this work.


\begin{thebibliography}{99}
\bibitem{Peres} A. Peres, Phys. Rev. Lett. {\bf 77}, 1413 (1996).
\bibitem{Horodecki} M. Horodecki, P. Horodecki \& R. Horodecki, Phys. Lett. A, {\bf 223}, 1 (1996).
\bibitem{formation}V. Vedral, M.B. Plenio, M.A. Rippin \& P.L. Knight, Phys. Rev. Lett. {\bf 78}, 002275 (1997).
\bibitem{lognegativity} G. Vidal \& R.F. Werner, Phys. Rev. A, {\bf 65}, 032314 (2002).
\bibitem{hightempent} V. Vedral, N. J. Phys. {\bf 6}, 102 (2004).
\bibitem{heatcapent} M. Wiesniak, V. Vedral \& \v{C}. Brukner {\it Pre-print} quant-ph/0508193 (2005).
\bibitem{ghosh} S. Ghosh, T. Rosenbaum, G. Aeppli \& S. Coppersmith, Nature, {\bf 425} 48 (2003).
\bibitem{Vedral1} V. Vedral, Nature, {\bf 425} 28 (2003).
\bibitem{Dunning05} C. Dunning, J. Links \& H. Zhou, Phys. Rev. Lett. {\bf 94}, 227002 (2005).
\bibitem{Latorre04} J. I. Latorre, E. Rico \& G. Vidal, Quant. Inf. Comput {\bf 4}, 48 (2004).
\bibitem{anders} J. Anders, D. Kaszlikowski, C. Lunkes, T. Ohshima \& V. Vedral,  New J. Phys. {\bf 8} 140 (2006).
\bibitem{Narnhofer03} H. Narnhofer, Phys. Lett. A, {\bf 310}, 423 (2003).
\bibitem{simon} R. Simon, Phys. Rev. Lett. {\bf 84}, 2726 (2000).
\bibitem{Anders03} J. Anders, {\it Diploma Thesis} quant-ph/0610263 (University of Potsdam, 2003).
\bibitem{Braunstein05} S. Braunstein \& P. van Loock, Rev. Mod. Phys. {\bf 77} 513 (2005).
\bibitem{Adesso07} G. Adesso \& F. Illuminati, {\it Pre-print} quant-ph/0701221 (2007).
\bibitem{Loudon73} R. Loudon {\it The Quantum Theory of Light, 3rd Ed.}, (Oxford Scientific Publications, 2000).  
\bibitem{Williamson36} J. Williamson, Am. J. Math. {\bf 58}, 141 (1936).
\bibitem{Simon87} R. Simon, E. C. G. Sudarshan \& N. Mukunda, Phys. Rev. A {\bf 36}, 3868 (1987).
\bibitem{vlatkomarcellojacob} M. Terra Cunha, J. Dunningham \& V. Vedral, {\it Pre-print} quant-ph/0606149 (2006).
\bibitem{Ent2ndQuant} V. Vedral, C. Euro. JP, {\bf 2}, 289 (2003).
\bibitem{Heaney07} L. Heaney, J. Anders, D. Kaszlikowski \& V. Vedral, {\it Pre-print} quant-ph/0702067 (2007).
\bibitem{Yang62} O. Penrose \& L. Onsager, Phys. Rev. {\bf 104} 576 (1956); C. N. Yang, Rev. Mod. Phys. {\bf 34} 694 (1962). 
\bibitem{dagoextract} D. Kaszlikowski \& V. Vedral, {\it Pre-print} quant-ph/0606238 (2006).
\end{thebibliography}
\end{document}